\documentstyle[12pt,aaspp4,psfig]{article}

\lefthead{left}
\righthead{right}

\newcommand{\vvmax}{$V/V_{\rm max} \ $}
\newcommand{\avmax}{$\langle V/V_{\rm max}\rangle \ $}
\newcommand{\gtilde}
 {~ \raisebox{-1ex}{$\stackrel{\textstyle >}{\sim}$} ~}
\newcommand{\ltilde}
 {~ \raisebox{-1ex}{$\stackrel{\textstyle <}{\sim}$} ~}

\begin{document}

\small

\title{Does the Number Density of Elliptical Galaxies Change
at $z <$ 1?}

\author{Tomonori Totani$^1$ and Yuzuru Yoshii$^{2,3}$}

\altaffiltext{1}{Department of Physics, School of Science,
The University of Tokyo, Tokyo 113, Japan \\
e-mail: totani@utaphp2.phys.s.u-tokyo.ac.jp}

\altaffiltext{2}{Institute of Astronomy, Faculty of Science,
The University of Tokyo, 2-21-1 Osawa, Mitaka, Tokyo 181, Japan}

\altaffiltext{3}{Research Center for the Early Universe, Faculty of Science,
The University of Tokyo, Tokyo 113, Japan}

\begin{abstract}
%%%%%%%%%%%%%%%%%%% making the right hand side beautiful!
%\pretolerance=100
%\tolerance=200
%\rightskip=0pt

We have performed a detailed \vvmax test for a sample of the Canada-France
Redshift Survey (CFRS) for the purpose of examining whether the comoving
number density of field galaxies changes significantly at redshifts of
$z<1$.  Taking into account the luminosity evolution of galaxies which
depends on their morphological type through different history of star
formation, we obtain \avmax$\approx 0.5$ in the range of $0.3<z<0.8$, where
reliable redshifts were secured by spectroscopy of either absorption or
emission lines for the CFRS sample. This indicates that a picture of mild
evolution of field galaxies without significant mergers is consistent with
the CFRS data. Early-type galaxies, selected by their $(V-I)_{\rm AB}$
color, become unnaturally deficient in number at $z>0.8$
due to the selection bias, thereby causing a fictitious decrease of \avmax.
We therefore conclude that a reasonable choice of upper bound of redshift
$z\sim 0.8$ in the \vvmax test saves the picture of passive
evolution for field ellipticals in the CFRS sample, which was rejected
by Kauffman, Charlot, \& White (1996) without confining the
redshift range. However, about 10\%
of the CFRS sample consists of galaxies having colors much bluer than
predicted for irregular galaxies, and their \avmax is significantly larger
than 0.5.  We discuss this population of extremely blue galaxies in terms
of starburst that has just turned on at their observed redshifts.
\end{abstract}

\keywords{galaxies: elliptical and lenticular, cD --- 
galaxies: evolution --- galaxies: formation}

\section{Introduction}

The standard scenario of formation of elliptical galaxies is an initial 
starburst in dissipative gas collapse at very high redshift, followed by 
passive luminosity evolution to the present (Larson 1974; Tinsley \& Gunn 
1976).  A galaxy evolution model of stellar 
population synthesis based on such a 
scenario well reproduces the present spectral energy distribution (SED) of 
elliptical galaxies and naturally explains their color--magnitude relation 
owing to the galactic wind which stops the starburst (Arimoto \& Yoshii 
1987; Matteucci \& Tornamb\'{e} 1987).  An obvious consequence of
the standard scenario is that the comoving number density 
of elliptical galaxies does not change at redshifts of $z<1$.  
However, Kauffmann, 
Charlot, and White (1996, hereafter KCW) recently performed a \vvmax test 
(Schmidt 1968) for a sample of ellipticals in the Canada-France Redshift 
Survey (CFRS, Lilly et al. 1995a), and reported a striking result that 
their number density should significantly decrease towards $z\sim 1$.  
Such a number evolution of field ellipticals obviously contradicts with 
the standard scenario and might be explained alternatively by mergers of 
smaller stellar systems and/or gaseous disks (Toomre \& Toomre 1972; 
Kauffmann, White, \& Guiderdoni 1993; Baugh, Cole, \& Frenk 1996).

In KCW's analysis, elliptical galaxies are selected from the CFRS sample 
if their observed $(V-I)_{\rm AB}$ colors are redder than a color boundary 
which separates ellipticals from other morphological types in the 
$(V-I)_{\rm AB}-z$ diagram.  They placed the boundary using a 
Bruzual--Charlot model of population synthesis with 0.1 Gyr single 
starburst and 50 \% solar metallicity.  This boundary certainly gives 
a reasonable fraction of ellipticals which agrees with {\it local} galaxy 
surveys.   However, the choice of burst duration and metallicity is only 
{\it ad hoc} without any physical basis, therefore rendering some doubts 
as to whether the KCW's color boundary works successfully up to $z\sim 1$.

On the other hand, local galaxies are known to have different colors from
earlier to later types along the Hubble sequence, and this color difference 
is attributed to type-dependent variation of star formation history in 
galaxies (Tinsley 1980; Kennicutt, Tamblyn, \& Congdon 1994).  Therefore,
the evolution models which reproduce the present colors would give a more 
natural way to distinguish various types in the $(V-I)_{\rm AB}-z$ diagram.
In this Letter, using the type-dependent evolution models of galaxies 
developed by Arimoto \& Yoshii (1987, hereafter AY) and Arimoto, Yoshii, 
\& Takahara (1992, hereafter AYT), we perform a \vvmax test for the whole 
CFRS sample consisting of various types as a mixture.  

\section{Color--Redshift Diagram}

Figure \ref{fig:color-z} shows the observed $(V-I)_{AB}$ colors versus 
secure redshifts of 591 galaxies in the CFRS sample.  Also shown are 
the color trajectories based on the AY (10$^{12}M_\odot$) model for E/S0 
(solid line) and the AYT (I1) models for Sab (short-dashed), Sbc 
(long-dashed), Scd (dot-short-dashed), and Sdm (dot-long-dashed).
Here, the symbol I1 for spirals stands for the infall model with $n=1$, 
where $n$ is a power index of the Schmidt law (see AYT for details).  
For comparison, KCW's model used as the color 
boundary between E/S0 and other types is shown by the dotted line in
this figure.  

Unless otherwise stated, we use the $\Lambda$-dominated universe with 
$(h,\Omega_M,\Omega_\Lambda) =(0.7,0.2,0.8)$, motivated by the fact that 
the luminosity density evolution based on the AY and AYT models is 
consistent with the CFRS data only in this universe (Totani, Yoshii, \& 
Sato 1997, hereafter TYS).  For the redshift of galaxy formation we adopt 
$z_F=5$, because the I1 models with this value of $z_F$ well reproduce 
Madau et al.'s (1996, 1997) observation of the UV luminosity density at 
$z\sim 2-4$, as shown by TYS and Totani (1997).  Furthermore, as far as 
$z_F \gtilde 3$, 
the effect of changing $z_F$ is very small in interpreting the 
CFRS data which extend only to $z\sim 1.3$.  

We see from Fig.1 that the color trajectories well cover the distribution 
of the CFRS data, and this indicates an overall validity of the E/S0--Sdm
models used here. In order to incorporate the effect of luminosity evolution 
in the \vvmax test, we associate each of the 591 galaxies in the CFRS sample 
with a luminosity evolution obtained by interpolating between adjacent two
of the five E/S0--Sdm models which have the nearest 
$(V-I)_{AB}$ color to a sample.  
We simply assign the E/S0 or Sdm model to red or blue 
galaxies, respectively, which locate outside the region enclosed by the 
E/S0 and Sdm color trajectories. 

\section{The \vvmax Test}

For a galaxy of apparent magnitude $m$ and redshift $z$ in the sample, 
we define two volumes as
\begin{equation}
V = \int_{\max (z_l, z_{\rm min}) }^{z}\frac{dV}{dz}dz  \;\;\;
{\rm and} \;\;\;
V_{\rm max} = \int_{\max (z_l, z_{\rm min}) }^{\min (z_u, z_{\rm max})}
\frac{dV}{dz}dz  \;\; ,
\end{equation}
where the comoving volume element $dV/dz$ is a function of $z$ which 
depends on the cosmological parameters of $h$, $\Omega_M$, and
$\Omega_\Lambda$.  Here, $z_{\min}$ or $z_{\max}$ is the redshift at 
which the galaxy in question having the absolute magnitude $M$ would be 
observed at the bright ($m_{\min}$) or faint ($m_{\max}$) magnitude limit, 
respectively.  The CFRS is limited in apparent magnitude between 
$m_{\min}=17.5$ and $m_{\max}=22.5$ in the $I_{AB}$ band.  
The redshift $z_x$ for either $x$=`min' or $x$=`max' should satisfy 
the following equation:
\begin{equation}
m-m_{x}=\{K(z)+E(z)\}-\{K(z_x)+E(z_x)\}+5\log \frac{d_L(z)}{d_L(z_x)} \;\; ,
\end{equation}
where $m$ and $z$ without suffix are those of observed quantities, and 
$d_L$ is the luminosity distance.  The $K$-correction $K$ and evolutionary 
correction $E$ are calculated using the AY and AYT models for the galaxy
type assigned in \S 2.

Although the CFRS extends to $z=1.3$, we have to limit the redshift range, 
$z_l<z<z_u$, over which the determination of redshifts should be free from 
biases.  Following Lilly et al. (1995b), we here adopt ($z_l, z_u$) =(0.3, 1) 
and calculate the ratio \vvmax for the CFRS galaxies in this redshift range.
In practice, if the hypothesis of mild evolution of galaxies is correct, 
their values of \vvmax are expected to distribute uniformly between 
0 and 1 with the average of \vvmax equal to 0.5, i.e., 
\avmax=0.5, irrespective of thier type.

Our \vvmax statistics give \avmax=0.405 (E/S0), 
0.396 (Sab), 0.507 (Sbc), 0.561 (Scd), and 0.642 (Sdm) with their 1 $\sigma$ 
statistical error of 0.030.   A clear trend of increasing \avmax from 
earlier to later types is seen, as claimed by Lilly et al (1995b) from 
their \vvmax test without considering the effect of galaxy evolution.  
The significantly small values of \avmax below 0.5 for E/S0 and Sab also 
seem to be consistent with KCW's result.   We here note that all the CFRS 
galaxies have been divided into five different types of roughly equal 
numbers according to their $(V-I)_{AB}$ colors.  Our type mixture slightly 
differs from local surveys, but this difference hardly changes the result
throughout this paper.

Since the \vvmax test is known to give a biased result unless the sample 
is complete (Schmidt 1968; Avni \& Bahcall 1980), we should be careful in 
interpreting the CFRS data because of difficulties of redshift 
identification for early type galaxies at $z\gtilde 1$. 
The spectral range used for the CFRS covers 4250--8500 \AA, and there are 
basically no features which are useful for redshift identification below 
3727 {\AA} at rest for emission-line galaxies and 4000 {\AA} at rest for 
absorption-line galaxies.  Consequently, the redshift identification is 
difficult at $z\gtilde 1$, especially for early type galaxies which show 
the absorption lines only (Hammer et al. 1995; Crampton et al. 1995).  
We clearly see from Fig. \ref{fig:color-z} that the red, early type 
galaxies are unnaturally sparse at $z\gtilde 0.8$, while bluer galaxies 
extend beyond $z\sim 1$.  It is therefore important to examine how the
\vvmax test is affected by changing $z_u$ in the analysis. 

Figure \ref{fig:VVmax-z_u} shows \avmax as a function of $z_u$ for 
five different types.  A steady decline of \avmax with increasing $z_u$ 
is seen beyond $z_u\sim 0.8$ for early type galaxies due solely to their
paucity in the range of $z\gtilde 0.8$. In fact, the CFRS group has also
pointed out that the apparent absence of red quiescent objects beyond
$z=0.8$ is likely due to selection effects (see \S 7 of Hammer et al.
1997), which is very well consistent with our results.
Therefore we have to decrease $z_u$ down to 0.8 in the case of
performing the \vvmax test free from the selection bias.  
In low redshifts of $z \ltilde 0.3$, on the other hand, the decrease of
the CCD efficiency also makes it difficult to indentify red absorption
galaxies (Hammer et al. 1995; Crampton et al. 1995).
We have performed analysis varing $z_l$, and found that changing $z_l$
leads no appreciable effect on \avmax.
Therefore we use the fixed value of $z_l=0.3$ throughout.

Our \vvmax statistics with $(z_l, z_u)=(0.3, 0.8)$ 
give \avmax=0.478 (E/S0), 0.496 (Sab), 
0.518 (Sbc), 0.513 (Scd), and 0.583 (Sdm) with their 1$\sigma$ statistical 
error of 0.035.  These \avmax values except for Sdm are consistent with
0.5 within the statistical uncertainties.
In order to see the reason why the \avmax for Sdm is 
significantly larger than 0.5, we show the CFRS galaxies classified as Sdm 
in Fig. \ref{fig:EBG}, where \vvmax is plotted against $\Delta (V-I) 
[\equiv (V-I)_{AB}^{\rm obs} - (V-I)_{AB}^{\rm Sdm-model}]$.
This figure indicates that much bluer galaxies compared to the 
Sdm model have larger \vvmax, while galaxies which have the color 
nearly equal to the Sdm model have a relatively uniform distribution 
in \vvmax.  Apparently the large \avmax for Sdm is originated from
such extremely blue galaxies (EBGs) with $\Delta (V-I)<-0.1$ as shown
by the encircled points in Figs. \ref{fig:color-z} and \ref{fig:EBG}.
The number of EBGs is 43 in the redshift range of $0<z<1.3$, and constitutes 
about 10 \% of the 591 galaxies in the whole CFRS sample.  When the EBGs are 
removed, \avmax decreases from 0.583 to 0.547 for Sdm. The \avmax values 
still seem to increase from earlier to later types, and this might reflect 
a type transformation which is not included in our hypothesis of mild 
evolution. However, we avoid any decisive conclusion about this
possibility of type transformation, because the \avmax values for all the 
types except for EBGs are consistent with 0.5 within the statistical 
uncertainties.  Thereby we conclude that the hypothesis of mild evolution 
without significant mergers is well consistent with 
the CFRS data excluding the EBGs, in the redshift range of $0.3<z<0.8$. 

In the above analysis we have assumed the $\Lambda$-dominated universe
with $(h, \Omega_M, \Omega_\Lambda)=(0.7, 0.2, 0.8)$.  Here we repeat 
the calculations using the Einstein-de Sitter (EdS) universe with 
$(h, \Omega_M, \Omega_\Lambda)=(0.5, 1, 0)$ and the low-density, open 
universe with $(h, \Omega_M, \Omega_\Lambda) = (0.6, 0.2, 0)$.  
The results for three universe models are tabulated in Table 1, provided
that the EBGs are removed from the sample.  The average \avmax increases 
with increasing $\Omega_M$ or decreasing $\Omega_\Lambda$, and \avmax of 
elliptical galaxies is consistent with 0.5 in all the three universe models.
The \vvmax of E/S0 in the EdS universe, 0.519, should be compared to
0.410 of KCW.
The average \avmax for the composite of all types is consistent with 0.5 
in the $\Lambda$-dominated universe, and our hypothesis of mild evolution 
favors this universe.  We furthermore repeat the calculations using the 
no-evolution model of galaxies (i.e., only $K$-correction is included),
and the result is shown in the table for
the purpose of comparison.  It is evident that galaxy evolution
is only a small effect in the \vvmax test as far as
$z<1$, and our results are hardly affected by uncertainties of 
the galaxy evolution model.

\section{Discussion and Conclusions}

In addition to the selection bias in the CFRS mentioned in the previous
section, our use of AY model for E/S0 is also responsible for the conclusion 
presented here which is in sharp contrast with KCW's conclusion.  
The behavior of KCW's evolution model at $z<0.6$ is similar to the AY model,
except for $\sim$ 
0.5 mag blueward shift probably due to their choice of a constant
value of lower metallicity for all stellar populations.  It should be 
noted that an average of absolute magnitude for the CFRS galaxies of E/S0
type is $M_{I_{AB}} = -21.9$, after corrected to $z=0$, which obviously
corresponds to the AY $10^{11-12}M_\odot$ (baryon) models with the
luminosity-weighted average of stellar metallicities equal to $130-180\%$
of the solar (see Table 3 of AY).  On the other hand, KCW's use of much 
lower metallicity of $30-70\%$ of the solar is equivalent to using the
AY $10^{9-10}M_\odot$ (baryon) model with 
$M_{I_{AB}}\sim -17$ which corresponds
to much smaller E/S0 galaxies not observed in the CFRS.  This suggests that
KCW's evolution 
model is inappropriate to place a color boundary between E/S0 and
other types for the CFRS sample.  

In fact, KCW have also noticed the selection bias against early-type 
galaxies at redshifts close to unity in the CFRS sample.  Since the 
$(V-I)_{AB}$ colors are available for the CFRS galaxies with no redshift 
identifications, KCW evaluated a maximum redshift for which each 
unidentified galaxy would still lie above the KCW's curve in Fig. 1 
(the dotted line) and be classified as early type.  They performed the 
\vvmax test including the unidentified galaxies with these maximum 
redshifts and obtained \avmax = 0.451 which is still smaller than 0.5.  
It should be noted that KCW's estimte of the maximum redshifts is
heavily based on the steep rise of their model at $z \gtilde 0.6$ in Fig. 1.  
In contrast, our models of E/S0 and Sab, which are considered to be more 
appropriate for analyzing the CFRS sample, do not show such a steep rise 
probably because of the longer duration of starbursts (0.7Gyr compared
to 0.1 Gyr in KCW) and 
hence it is difficult or even impossible to define the maximum redshift for 
the unidentified, red galaxies.  Therefore, we consider that the estimate
of the maximum redshifts is highly uncertain, and the 
incompleteness should be avoided by performing the \vvmax test only in the 
range in which spectroscopic redshifts are secured with confidence.

\begin{figure}
  \begin{center}
    \leavevmode\psfig{file=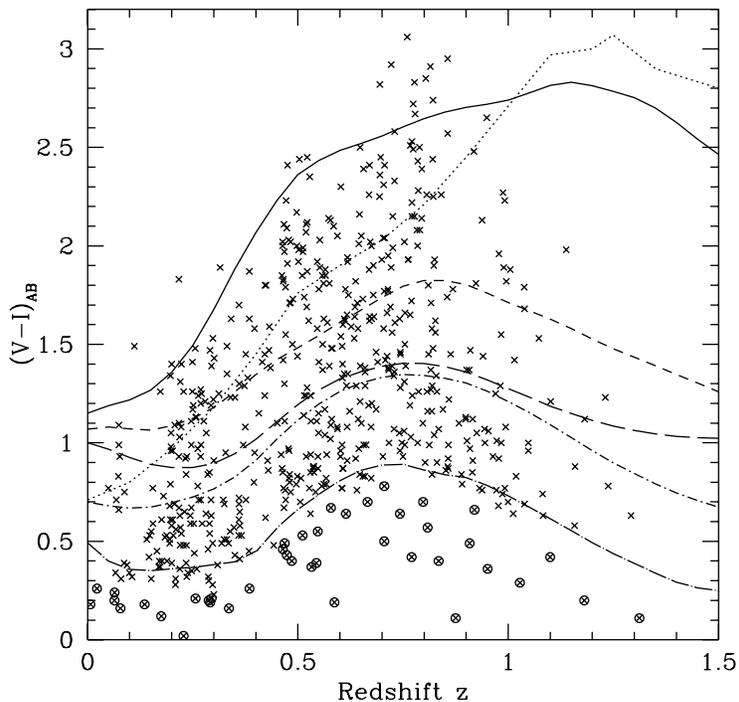,width=10cm}
  \end{center}
\caption{
The observed redshifts versus $(V-I)_{AB}$ colors of the galaxies in
the CFRS sample (Lilly et al. 1995a). The model curves are the galaxy 
evolution models of E/S0 (solid), Sab (short-dashed), Sbc (long-dashed),
Scd (dot-short-dashed), and Sdm (dot-long-dashed).  The $\Lambda$-dominated 
universe with $(h, \Omega_M, \Omega_\Lambda)=(0.7, 0.2, 0.8)$ and 
the formation redshift of $z_F = 5$ are assumed.  The dotted line is
the model used by Kauffmann, Charlot, \& White (1996).  The encircled
data points are the extremely blue galaxies (EBGs) whose colors are bluer 
than the Sdm curve by more than 0.1 mag (see text for detail).
}
\label{fig:color-z}
\end{figure}

\begin{figure}
  \begin{center}
    \leavevmode\psfig{file=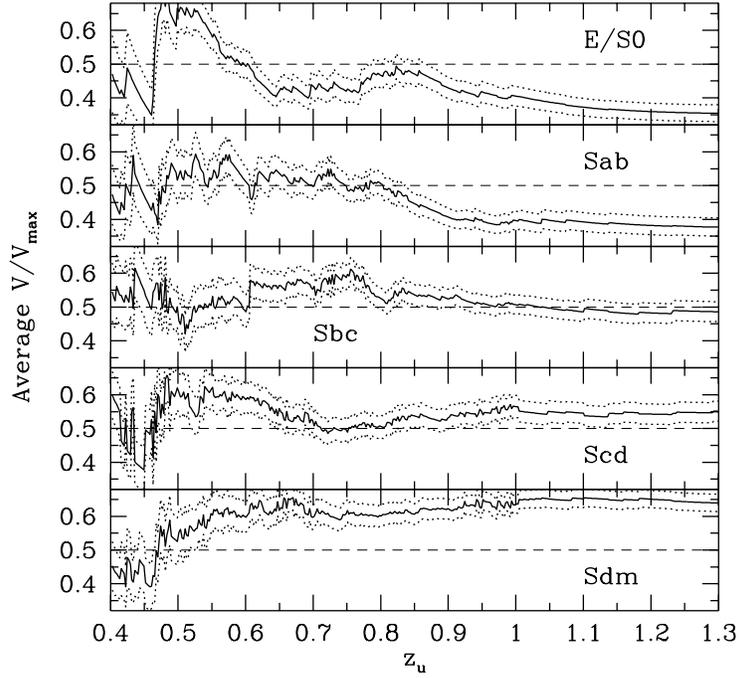,width=10cm}
  \end{center}
\caption{The average \avmax for five different types of galaxies as a 
function of the upper redshift bound $z_u$ with the fixed lower bound of 
$z_l=0.3$. The solid and dotted lines represent \avmax and the 1 $\sigma$ 
statistical error, respectively.  The $\Lambda$-dominated universe with 
$(h, \Omega_M, \Omega_\Lambda) = (0.7, 0.2, 0.8)$ and the formation 
redshift of $z_F = 5$ are assumed. If \avmax is 0.5 (dashed line), the 
comoving number density of galaxies is conserved in the specified range
of $z_l<z<z_u$.
}
\label{fig:VVmax-z_u}
\end{figure}

\begin{figure}
  \begin{center}
    \leavevmode\psfig{file=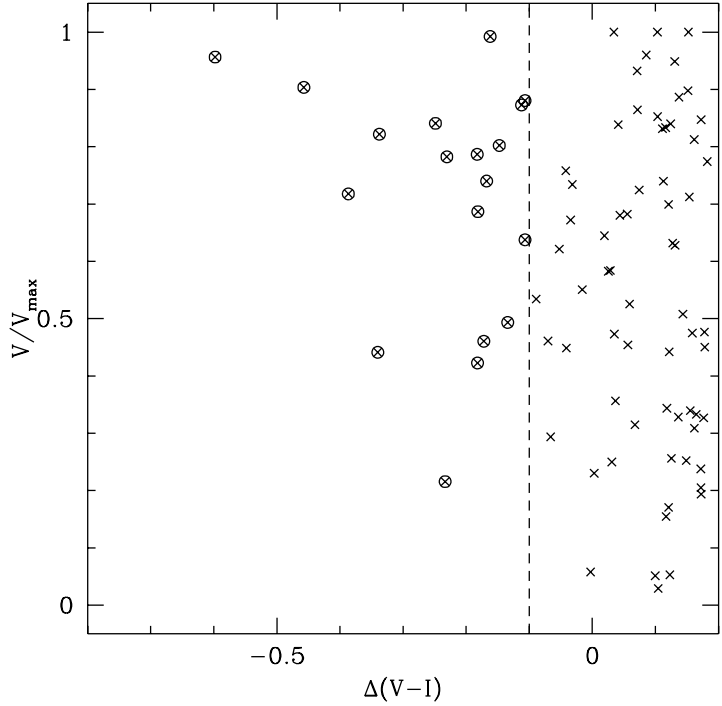,width=10cm}
  \end{center}
\caption{The \vvmax versus $\Delta (V-I)$ plot for the Sdm galaxies 
selected from the the CFRS sample by their colors, where 
$\Delta (V-I) \equiv (V-I)_{AB}^{\rm obs} - (V-I)_{AB}^{\rm Sdm-model}$. 
The trend of larger \vvmax for much smaller $\Delta (V-I)$ can be seen.
The encircled data points are the galaxies with $\Delta (V-I)<-0.1$, 
which we call extremely blue galaxies (EBGs).  The $\Lambda$-dominated 
universe with $(h, \Omega_M, \Omega_\Lambda) = (0.7, 0.2, 0.8)$ and the 
formation redshift of $z_F = 5$ are assumed.
}
\label{fig:EBG}
\end{figure}

Let us discuss more about the nature of EBGs, which are distributed widely 
in the redshift range of $0<z<1$ (Fig. \ref{fig:color-z}).  The fact
that their \avmax value is larger than 0.5 suggests that the EBGs, soon 
after observed at $z$, must disappear or change their color to be 
classified as redder galaxies.  A likely mechanism for such sudden 
fading or reddening is that the EBGs are being observed during intensive 
starburst as inferred from their extremely blue colors.  In either cases
the evolution of the EBGs may not be predicted
by the usual Sdm models including 
ours, in which  the star formation rate 
is nearly constant to make the present colors as blue as local Sdm 
galaxies.  However, if the star formation is suddenly stopped on the way, 
the colors turn considerably redder than predicted by the Sdm model.
It is therefore tempting to examine whether such starburst models for the 
EBGs lead to \avmax$\sim 0.5$ for the whole CFRS galaxies. 

In this letter, we have performed a detailed \vvmax test for the CFRS 
sample taking into account the AY and AYT models of galaxy evolution. 
Our hypothesis that all field galaxies evolve mildly according to 
type-dependent star formation histories without significant mergers is 
consistent with the CFRS data in $0.3<z<0.8$ excluding EBGs, especially in the 
$\Lambda$-dominated universe.
Consequently, the rejetion of passive evolution of ellipticals claimed
by KCW is not valid. The CFRS data do not allow us to discriminate
between the passive evolution and the number evolution of ellipticals
claimed by KCW.

If we parametrize the evolution
of comoving number density of E/S0 galaxies as $\propto (1+z)^\gamma$,
then the analysis in the limited redshift range of $0.3<z<0.8$ gives
$\gamma_{\rm E/S0}=0.6\pm 1.7$, $-0.1\pm 1.7$, and $-0.8\pm 1.7 $ for
the EdS, open, and $\Lambda$-dominated universe models, respectively.
The errors are larger than KCW's errors, because the redshift range is
limited in our analysis.  For all the CFRS galaxies excluding the EBGs, the
analysis within the same redshift range gives $\gamma_{\rm all}=1.8\pm 0.7$,
$1.1\pm 0.7$, and $0.5\pm 0.7$ in the same order as above.
These $\gamma$ values, though not so significant as claimed by KCW,
would impose an important constraint on the number evolution of field
galaxies in the framework of the hierarchical clustering of galaxies in
the universe.

This work has been supported in part by the Grant-in-Aid for Scientific 
Research (3730) and Center-of-Excellence (COE) research (07CE2002) of the 
Ministry of Education, Science, and Culture of Japan. TT has been 
supported by JSPS Research Fellowships for Young Scientists.

%%%%%%%%%%%%%%%%%%%% TABLES %%%%%%%%%%%%%%%%%%%%%%%%%%%%%%%%%%%
\begin{table}
\scriptsize
%\footnotesize
\begin{center}
\caption{The \vvmax Statistics}
\begin{tabular}{cccccccc} 
\hline  \hline Cosmology
& Galaxy & \multicolumn{6}{c}{\avmax \tablenotemark{\it a}} \\
\cline{3-8}
$(h, \Omega_M, \Omega_\Lambda)$ &  Evolution
&  E/S0  &  Sab  &  Sbc  &  Scd  &  Sdm  &  All Types  \\
\hline
(0.5, 1.0, 0.0) & on &
.519 $\pm$     .038 & 
.532 $\pm$     .034 & 
.540 $\pm$     .034 & 
.550 $\pm$     .034 & 
.569 $\pm$     .033 & 
.542 $\pm$     .016   \\

(0.6, 0.2, 0.0)  & on &
.494 $\pm$     .038 & 
.508 $\pm$     .034 & 
.530 $\pm$     .033 & 
.524 $\pm$     .033 & 
.575 $\pm$     .034 & 
.526 $\pm$     .016   \\

(0.7, 0.2, 0.8) & on &
.478 $\pm$     .039 & 
.496 $\pm$     .036 & 
.518 $\pm$     .034 & 
.513 $\pm$     .033 & 
.547 $\pm$     .033 & 
.511 $\pm$     .016   \\

(0.7, 0.2, 0.8) & off &
.471 $\pm$     .038 & 
.501 $\pm$     .038 & 
.537 $\pm$     .033 & 
.536 $\pm$     .031 & 
.539 $\pm$     .037 & 
.517 $\pm$     .016  \\
\hline \hline
\end{tabular}
\end{center}
\tablenotetext{a}{
Galaxy types used here are based on the color-aided classification
(\S 2) and are slightly different from the local surveys.
The EBGs are not included in the statistics (see text for detail).}

\end{table}

%%%%%%%%%%%%%%%%%%%% REFERENCES %%%%%%%%%%%%%%%%%%%%%%%%%%%%%%

%%%%%%%%%%%%%%%%%%%%%% FIGURES %%%%%%%%%%%%%%%%%%%%%%%%%%%%

\newpage

\end{document}